# Zero-bias conductance peak observed in Au/$FeSe_{0.3}Te_{0.7}$ ramp-type junctions and its implication on the superconducting order parameter

Y.S. Li[1], J.Y. Luo[1], M.J. Wang[2], T.J. Chen[2], M.K. Wu[1,3], C.C. Chi[3]
[1]*Institute of Physics, Academia Sinica, Nankang, Taipei 115, Taiwan*
[2]*Institute of Astrophysics and Astronomy, Academia Sinica, Taipei 115, Taiwan*
[3]*Department of Physics, National Tsing Hua University, Hsinchu 300, Taiwan*
E-mail: cchi@phys.nthu.edu.tw

## ABSTRACT

Based upon the currently predominant theory of Fe-based superconductors, the order parameter symmetry is the nodaless s±-wave. Andreev conductance involving normal metals and such superconductors does not have a zero bias conductance peak (ZBCP) except maybe for a very narrow range of a junction-dependent fitting parameter. We have measured differential conductance of the in-plane boundary junctions between Au and superconducting $FeSe_{0.3}Te_{0.7}$ films. The Au/$FeSe_{0.3}Te_{0.7}$ ramp-type junctions are fabricated with the bias current direction along the <110> axis of the epitaxial $FeSe_{0.3}Te_{0.7}$ film. We have always observed a pronounced ZBCP as well as some gap-like features in the conductance spectra of all samples studied, thus our experimental results suggest not the s±-wave theory but the possible existence of a nodal point around 45° in at least one of the gaps in $FeSe_{0.3}Te_{0.7}$ superconductor.

## I. INTRODUCTION

The discovery of the iron-based superconductor (Fe-SC) [1-7] has opened up a new class of superconducting materials, arousing extensive research on their physical properties in recent years. Among these new materials, iron selenide has the simplest structure, which contains only the FeSe sheet without any interlayer. Its critical temperature ($T_C$) is lowest at about 8K[8]. Partial substitution of Se with Te can increase the $T_C$ of this compound to a maximum of 14K [9-11].

One of the often asked questions about the iron-based superconductors is whether they are conventional superconductors or not. Early experimental results using point contact Andreev reflection (PCAR) data by T.Y. Chen et. al. [12] indicated a single BCS-like gap (nodaless) in $SmFeAsO_{0.85}F_{0.15}$. However, the later PCAR data obtained by other researchers show two nodaless gaps [13-16]. Furthermore, still other PCAR [17] and STM [18] data infer that at least one gap in Fe-SC is nodal because of the appearance of the zero bias conductance peak (ZBCP). On the theoretical side, different kinds of gap symmetries, such as p-wave[19], d-wave[20], and s±-wave[21-26], have been proposed. Recently more theoretical papers seem to support the s± symmetry, i.e. two s-wave order parameters, one for the electron-type and one for the hole-type of the Fermi surfaces, with a π phase difference between the two. Several recent experiments, using different techniques such as Josephson tunneling[27], phase sensitive SQUID measurement[28], and STM[29] measurement, all suggest the s± symmetry for Fe-SC.

However, most of the experimental data to date are obtained by using either poly-crystalline samples or c-axis tunneling with single crystal samples. To our knowledge, there is no experimental report about the in-plane tunneling of iron-based superconductors. In this paper, we report such transport measurements between the boundary of a normal metal and $FeSe_{0.3}Te_{0.7}$ along the <110> direction of the latter. We have observed the ZBCP in such boundaries, which seems to be inconsistent with the proximity effect of the s± model. In addition, several gap-like features are also observed, which are more or less consistent with other previously reported experimental results.

## II. JUNCTION FABRICATION

Using the capability of depositing high quality epitaxial $FeSe_{0.3}Te_{0.7}$ films developed in our lab, we fabricate $Au/FeSe_{0.3}Te_{0.7}$ ramp-type junctions to study the proximity effect. The geometry of our sample is shown in Fig. 1 (a) and (b). The tunneling direction is along <110> of $FeSe_{0.3}Te_{0.7}$ as indicated in Fig.1(a). The cross-section of junction is shown in Fig.1(b). Individual layers of 5nm NbTi, 100nm Au, 5nm NbTi and 100nm $SiO_2$ are sequentially sputtered onto (100)MgO substrate as shown in Fig. 1(b). The purpose of depositing the thin NbTi layers is to improve the adhesion between Au and MgO and between Au and $SiO_2$ interfaces. Standard photo-lithography technique is applied to the sample to define the first pattern labeled "Au", shown in Fig. 1(a), and Ar-ion milling is used to form the ramp, indicated in Fig. 1(b). After stripping the photo-resist, the $FeSe_{0.3}Te_{0.7}$ layer is deposited by using a pulsed laser (Lambda physik KrF 248nm) deposition technique. The substrate temperature is 350℃, target-to-substrate distance is 48mm, and laser energy density is 1~1.5J/cm$^2$. The stoichiometry of $FeSe_{0.3}Te_{0.7}$ (hereafter we designate this as FeSeTe) is a nominal composition. Subsequently, the photolithography technique and Ar-ion milling are used to produce our junctions, shown in Fig. 1(a) and (b). Finally, for making electrical contacts to the sample, the covered $SiO_2$ layer on the gold pad is removed by using another photolithography and Ar-ion milling process. Our junction is 10 μm wide and the resulting ramp angle with respect to the substrate plane is about 30°.

## III. RESULTS AND DISCUSSIONS

Figure 2(a) shows the θ-2θ x-ray diffraction pattern of our typical FeSeTe film. Since only the (00 *l*) peaks are present, our FeSeTe thin film is c-axis oriented. Figure 2(b) shows the R-T curves of the three junctions reported in this paper—$J_I$(black), $J_{II}$(red) and $J_{III}$(blue). The superconducting transition width ΔT(90%Rn-10%Rn) is about 1K. $J_{III}$ has gone through an additional in-situ annealing process (350℃, 30min) after depositing the FeSeTe film, while $J_I$ and $J_{II}$ have not. Figure 2(c)-(e) are the in-plane Φ-scan of the (101) peak of FeSeTe in the pad region of three junctions respectively. The film in-plane orientation is mainly <110>FeSeTe || <100>MgO with ΔΦ(FWHM) ~ 4°. The R-T curve of the FeSeTe film with additional annealing process shows a slightly higher transition temperature and a cleaner in-plane Φ-scan data.

Figure 3(a) shows the differential conductance spectra of three ramp-type junctions measured at 2K. All of the measured spectra exhibit the ZBCP. In addition, there are some gap-like features appearing in the spectra indicated by arrows as shown in (a) for $J_{II}$ and $J_{III}$, in (b) for $J_I$. In order to understand the behavior of the ZBCP and the gap-like features, we measure the temperature and magnetic field dependences of the conductance spectra of the three junctions.

Figure 4(a) shows the conductance spectra of $J_I$ at several temperatures between 2K to 10.5K. The ZBCP becomes smaller when the temperature approaches $T_C$. Note that the relatively steep drop of the conductance at higher temperatures and higher bias voltages shown in Fig.4(a) is due to the fact that part of the FeSeTe film becomes normal as the critical current density of the film is exceeded. In addition to the ZBCP, there are three gap-like features shown in Fig. 4(a). For the smallest one, we can only distinguish it near 2K, and when the temperature arises, this gap-like feature is overwhelmed by the ZBCP. For the larger gap like structures, they progressively become smaller as the temperature rises. In order to extract the position of these three gap-like features, we use the voltage values corresponding to the extreme values of the double derivative of the conductance spectrum (not shown here) as our practical criterion. If we take the derivative on the raw conductance data, the resulting derivative curve would be very noisy due to the high frequency noise present in the data. Thus, we use the Fourier filtering technique to remove the high frequency noise and take the double derivative on the noise-removed spectra to determine the voltage values. Figure 4(b) shows the temperature dependence of two larger gaps up to 7K, since the smallest one is obtained only at 2K. Three gap-like values $\Delta_1$(2K), $\Delta_2$(2K)and $\Delta_3$(2K) are 0.90meV, 2.66meV and 4.23meV respectively. When temperature is above 7K, no gap-like structures can be obtained from our analysis because the conductance spectrum becomes too smooth to have any visible gap features. From the RT measurement, we know that the $T_C$ of the junction is about 11K. On the other hand, the temperature dependence of the gap-like features shown in Fig. 4(b) is very different from the BCS trend with $T_C$ = 11K. On the other hand, the temperature dependence of the largest gap structure seems to be consistent with a BCS-like trend with a lower transition temperature about 7.7 K.

Figure4(c) and (d) show the magnetic field dependence of the conductance spectra of $J_I$ at 2K. The ZBCP decreases when the magnetic field is applied from 0 to 0.1T. Further increasing the strength of the magnetic field causes the ZBCP to become gradually smaller, but the ZBCP clearly survives up to 9T. In addition to the ZBCP, the three gap-like features are smeared when the magnetic field is applied. Just like the temperature dependence of the spectrum, the drop of the conductance spectrum at higher magnetic field and high bias region is due to exceeding the critical current density of the FeSeTe film.

We have also measured the temperature dependence of the conductance spectra of $J_{II}$ and $J_{III}$. After extracting the temperature dependence of the voltage positions of the gap-like structures of these two junctions, we summarize the results in the Fig. 5, including $J_I$. There are three gap-like features in $J_I$, two gap-like features in $J_{II}$ and one gap-like feature in $J_{III}$. From the RT measurement, we know the $T_C$ of all three junctions is about 11K. The temperature dependences of the gap-like features do not follow the BCS-like trend. The voltage position versus temperature of the middle gap-like structure of $J_I$ is very close to that of $J_{III}$. We note that this energy is consistent with other reports [30,31] for the same material, i.e. $FeSe_xTe_{1-x}$. For the largest gap-like structure in $J_I$, if we take $\Delta_0 = 4.23$meV and $T_C = 11$K, the corresponding coupling strength, $\Delta_0/ kT_C$, is 4.46, which is comparable to those reported for FeAs-type superconductors [13-15]. For the larger gap-like structure in $J_{II}$, if we take $\Delta_0 = 1.84$meV and $T_C = 11$K, the corresponding coupling strength, $\Delta_0/ kT_C$, is 1.94, which is similar to the coupling strength in FeAs-type superconductors found in some reports[12,13,16,18]. For the smallest gap-like structure in $J_I$ and smaller gap-like structure in $J_{II}$, if we take $\Delta_0 = 0.90$ and 1.02meV and $T_C = 11$K, the corresponding coupling strengths, $\Delta_0/ kT_C$, are 0.95 and 1.08. Coupling strengths of these magnitudes have also appeared in some reports for FeAs-type superconductors [16,17].

Although the gap-like features do show some variations from junction to junction, the ZBCP always appears in the spectra for all of our studied samples, including those not explicitly shown here. To our knowledge, there are three possible mechanisms to form the ZBCP: s-wave, d-wave, and currently favored s±-wave. For the s-wave scenario, the ZBCP forms only if the junction is in the clean contact limit. Our numerical conductance spectra based upon this possible mechanism, with parameters $\Delta(0) = 0.5$meV, Z = 0(clean contact), $T_C = 11$K and the BCS-like temperature dependence of $\Delta(T)$, are shown in Fig. 6. They are clearly different from our measured spectra in shape as well as the temperature dependence. Thus, we believe a simple s-wave pairing symmetry can be ruled out for FeSeTe.

The d-wave scenario can be ruled out by the Josephson tunneling experiment [27], in which the observed Josephson coupling along the c-axis between $Ba_{1-x}K_xFe_2As_2$ and a conventional superconductor suggests the existence of an s-wave symmetry in this class of iron pnictide superconductors. Furthermore the fact that the phase sensitive experiment [36], using the scanning SQUID microscopy to study the Nd-1111 sample, did not find any paramagnetic Meissner effect also seems to disfavor d-wave scenario.

The third possible mechanism to form ZBCP is s±-wave. T. Hanaguri *et al.*[29] has suggested the unconventional s-wave (s±-wave) in Fe(Se,Te). A. A. Golubov *et al.*[34]

modified the classical BTK theory by including the s±-wave scenario. We use this extended BTK model to calculate the conductance spectra. In this theory, there is a dimensionless parameter α which is defined by:

$$\alpha = \alpha_0 \frac{\phi_q(0)}{\phi_p(0)},$$

where $\alpha_0$ denotes the ratio of the probability amplitude for an electron from normal metal to tunnel into the second band over the probability amplitude to tunnel into the first band and it may actually change from contact to contact as the interface property change, $\phi(0)$ denotes the Bloch function at the junction interface in two-band metal and p and q denote the Fermi vector for the first and second band, respectively. The only way to form the ZBCP requires a specific condition:

$$\alpha^2 = \frac{\Delta_1}{\Delta_2}.$$

We simulate this situation and put $\Delta_1$=1meV and $\Delta_2$=2meV as suggested in Ref. [34]. The ZBCP indeed occurs when $\alpha = 1/\sqrt{2} \approx 0.707$ even at T = 2K. When α is a little larger or smaller than this specific value, the ZBCP either disappears or turns into finite voltage conductance peaks as shown in Fig. 7.

Figure7 shows the conductance spectra when α = 0.770, 0.714, 0.707, 0.7035, 0.7 and 0.63 with Z = 0.4 at 2K. When α = 0.700 or 0.714 we can clearly see two finite voltage conductance peaks. The value of α is only 1% different than $1/\sqrt{2} \approx 0.707$. Since we always observed the ZBCP in our conductance spectrum at 2K, then α should be nearly constant for all junctions. However, as pointed out by Golubov *et al.*[34], α should depend on the junction's interface properties. It is difficult to imagine that we can have such coincident values for each different junction. Thus we believe that s± model does not provide a natural explanation for our experimental results.

Finally, another possibility is that there are nodes in one of the gaps in superconducting FeSeTe. Recently, C. L. Song *et al.* [35] indeed proposed the existence of nodes based on their STM tunneling data of c-axis FeSe thin films. Since the tunneling direction of our junction is along 45° of the FeSeTe lattice, we speculate that there may exist a nodal point around 45° in one of the gaps in FeSeTe superconductor.

## IV. CONCLUSIONS

In summary, we study the temperature and magnetic field dependences of the conductance spectra of the Au/FeSeTe ramp-type junctions. All junctions show the ZBCP which survives in the presence of magnetic fields up to 9T. In addition to the ZBCP, we also observe the gap-like features in our spectra. The temperature dependences of the size of the gap-like features do not follow the BCS-like trend. When magnetic field is applied, these features become smeared. Although the gap-like features show some variation from junction to junction, the ZBCP always appears in the spectrum. According to the proximity theory with s± symmetry for superconducting gaps by Golubov *et al.*[34], the α parameter has to be a specific value to show ZBCP. Since we have observed ZBCP in all samples with different contact resistances, we believe that it is unlikely that s± symmetry is the correct theory for FeSeTe samples. Thus we favor the possibility that there are nodes in one of the gaps, which has recently been suggested by C. L. Song *et al.* [35]. Because the tunneling direction of our junctions is along the <110> FeSeTe lattice, we speculate that the nodal point is around 45° in one of the gaps in FeSeTe superconductor.

**ACKNOWLEDGMENTS**
This work was supported by National Science Council, Taiwan, R.O.C. Grant Numbers: NSC 99-2120-M-007-002 and NSC 99-2112-M-001-028-MY3.
*Electronic address: cchi@phys.nthu.edu.tw

**Figure Captions**

FIG. 1. Schematic layout of the ramp-type Au/FeSeTe junction: (a)top view and (b)cross-sectional view.

FIG. 2. (a) The θ-2θ x-ray diffraction pattern of our thin film. (b) The R-T curves of the three junctions $J_I$, $J_{II}$, $J_{III}$ reported here. The inset shows the detail of the R-T curves in the superconducting transition region. (c)-(e) In-plane Φ-scans of (101) peak of FeSeTe in the pad region of the three junctions respectively.

FIG. 3. (a) Conductance spectra of the three junctions measured at 2K. All spectra show the ZBCP. In addition to the ZBCP, there are some gap-like features in the spectrum as indicated by arrows shown in (a) for $J_{II}$ and $J_{III}$, in (b) for $J_I$.

FIG. 4. (a) Conductance spectra of the $J_I$ with temperature ranging from 2K to 10.5K. In addition to ZBCP, we can clearly see three gap-like features at 2K. The two large gap-like energies decrease with increasing temperature before they become indistinguishable in the background. (b) The temperature dependence of the voltage position of the two larger gap-like features. For practical purposes, the voltage position of the gap-like structures is determined by the extreme values of the double derivative curves of the measured conductance spectra. Three gap-like values $\Delta_1$, $\Delta_2$ and $\Delta_3$ at 2K are 0.90meV, 2.66meV and 4.23meV, respectively. (c) The magnetic field dependence, ranging from 0T to 0.9T, of the conductance spectra of the $J_I$ at 2K. (d) The magnetic field dependence, ranging from 1T to 9T, of the conductance spectra of the $J_I$ at 2K.

FIG. 5. Summary of the gap-like features of $J_I$, $J_{II}$ and $J_{III}$. $T_C$, indicated by the arrow, is determined by R-T data.

FIG. 6. Calculated conductance spectra using the s-wave scenario with T = 0.5, 2, 4, 6, 8 and 10K. The simulation parameters are: $\Delta(0)$ = 0.5meV, Z = 0 (clean contact), $T_C$ = 11K and $\Delta(T)$ is varied with BCS-like temperature dependence.

FIG. 7. Simulation results from the extended BTK theory [34]. The values of α = 0.770, 0.714, 0.707, 0.7035, 0.7 and 0.63 with Z = 0.4 at 2K.

Fig. 1

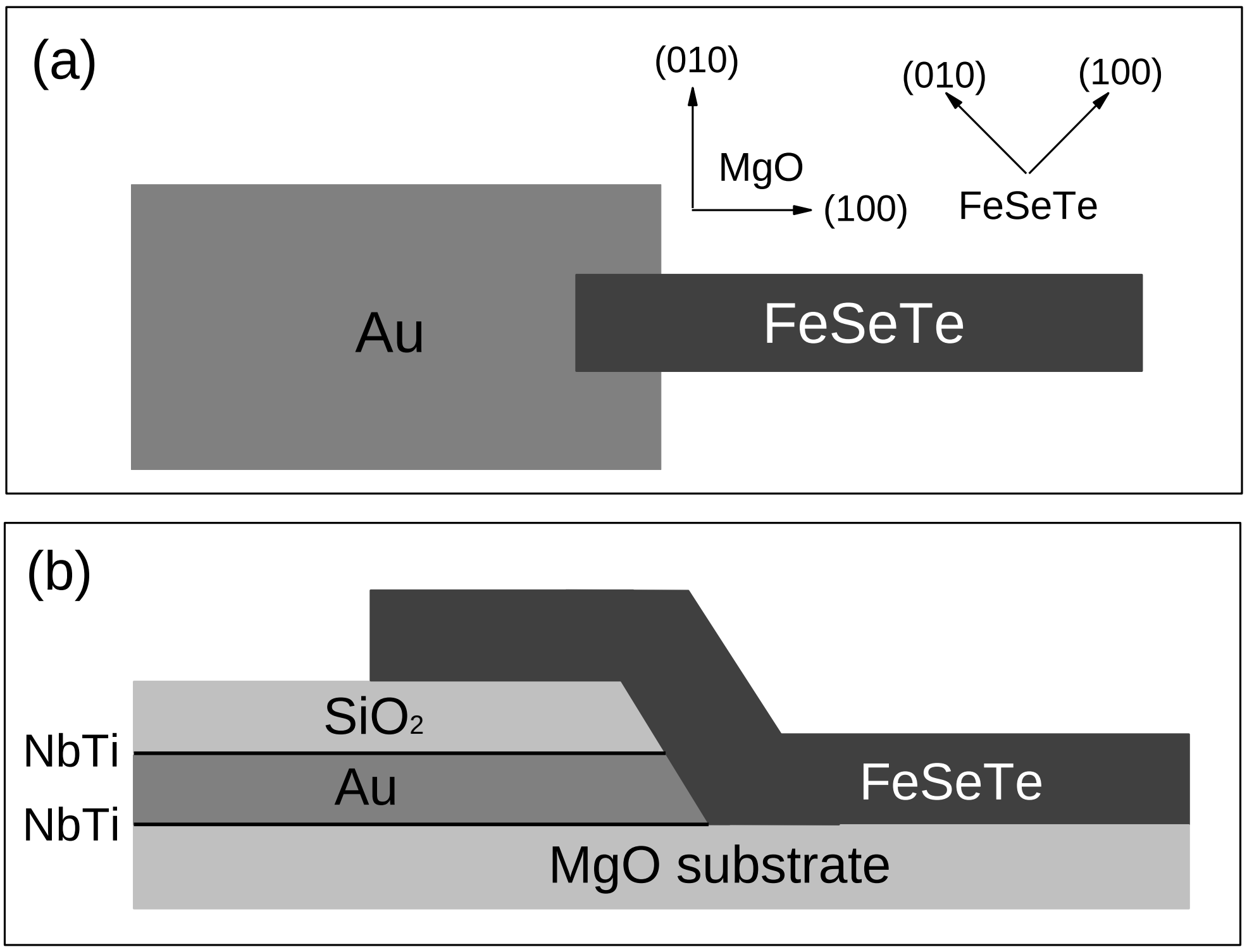

Fig.2.

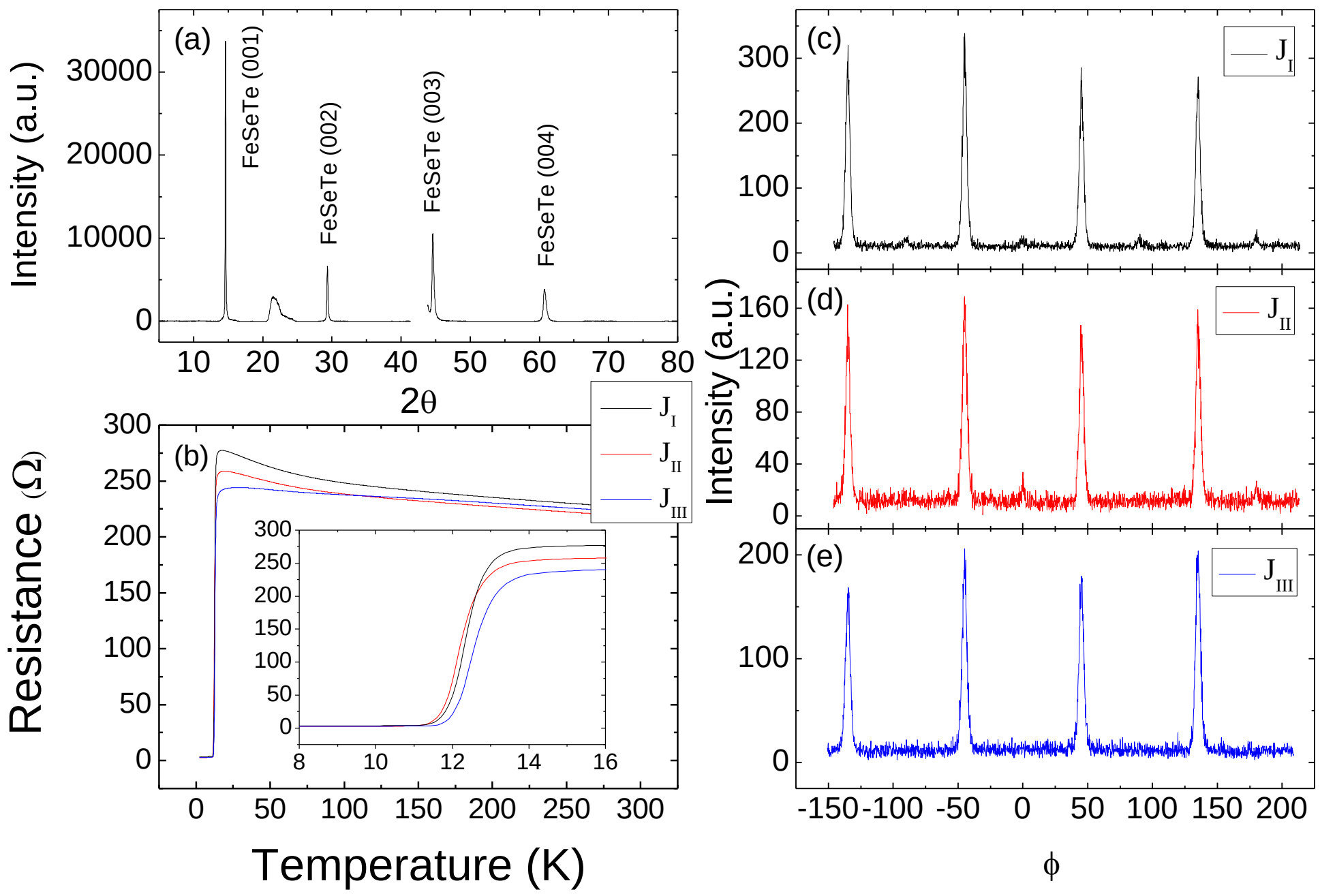

(a)
Intensity (a.u.)
30000
20000
10000
0
FeSeTe (001)
FeSeTe (002)
FeSeTe (003)
FeSeTe (004)
10 20 30 40 50 60 70 80
2θ
(b)
Resistance (Ω)
300
250
200
150
100
50
0
0 50 100 150 200 250 300
Temperature (K)
J_I
J_II
J_III
8 10 12 14 16
(c)
(d)
(e)
Intensity (a.u.)
-150 -100 -50 0 50 100 150 200
ϕ

Fig. 3.

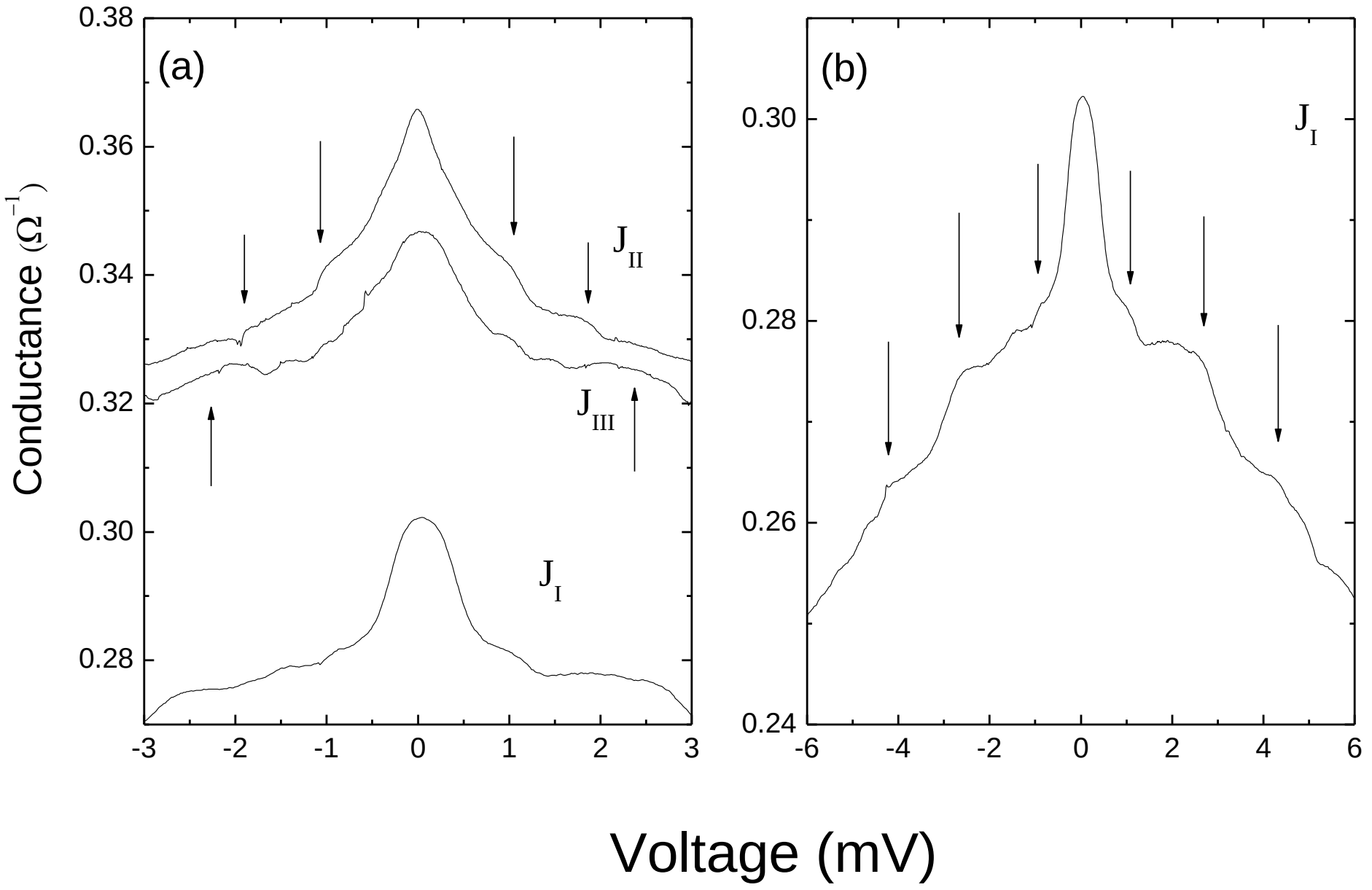

Fig. 4.

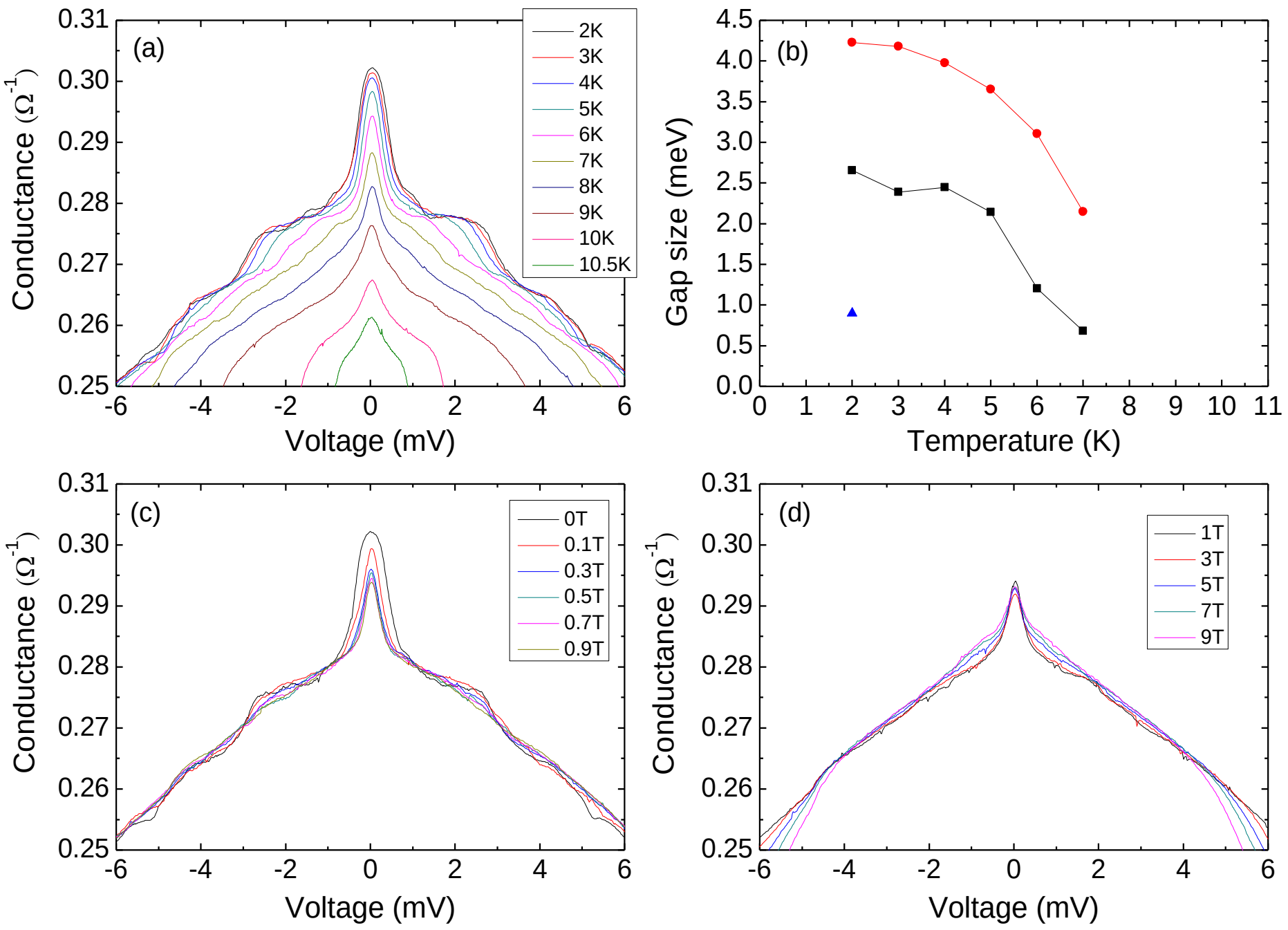

Fig. 5.

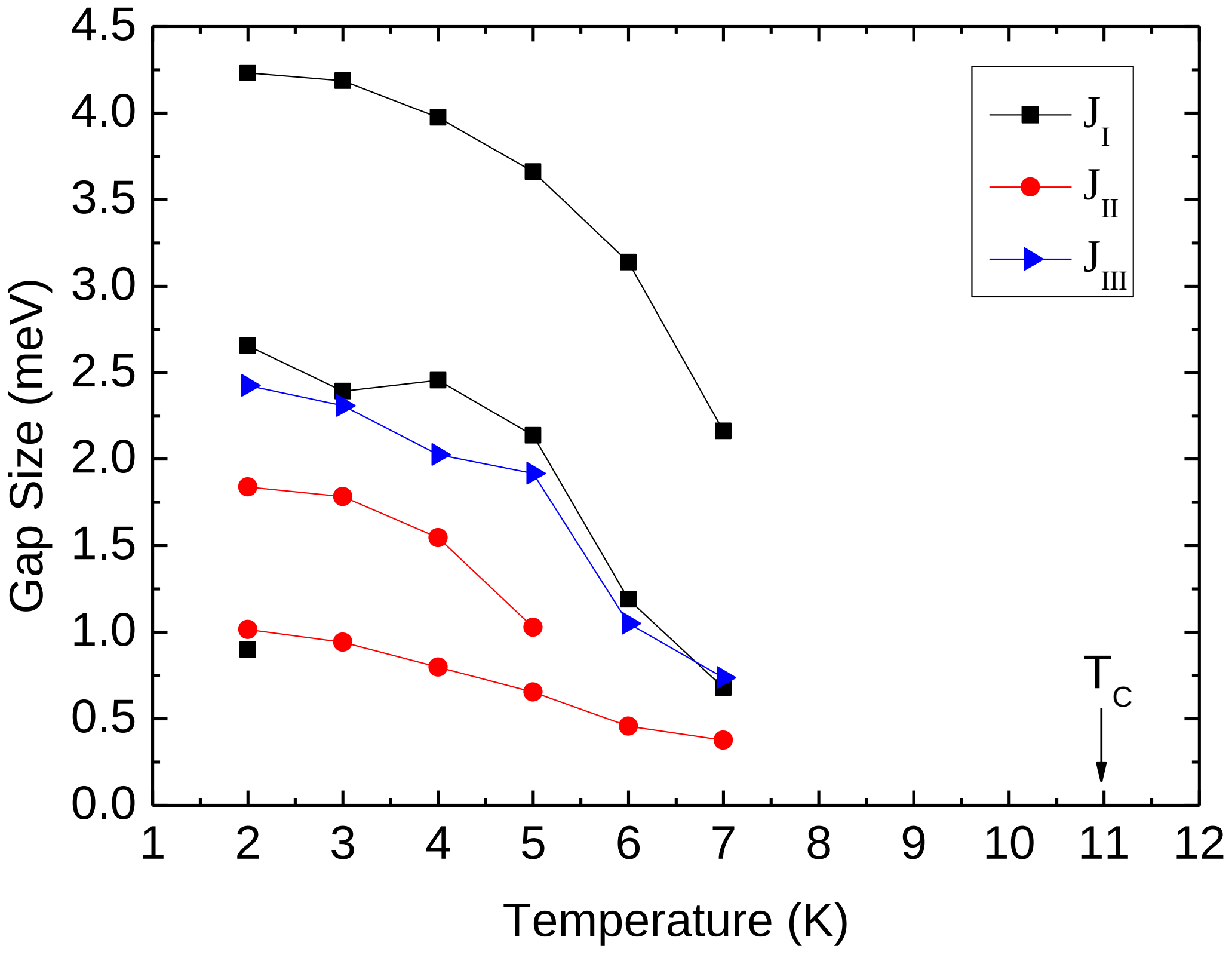

Fig. 6.

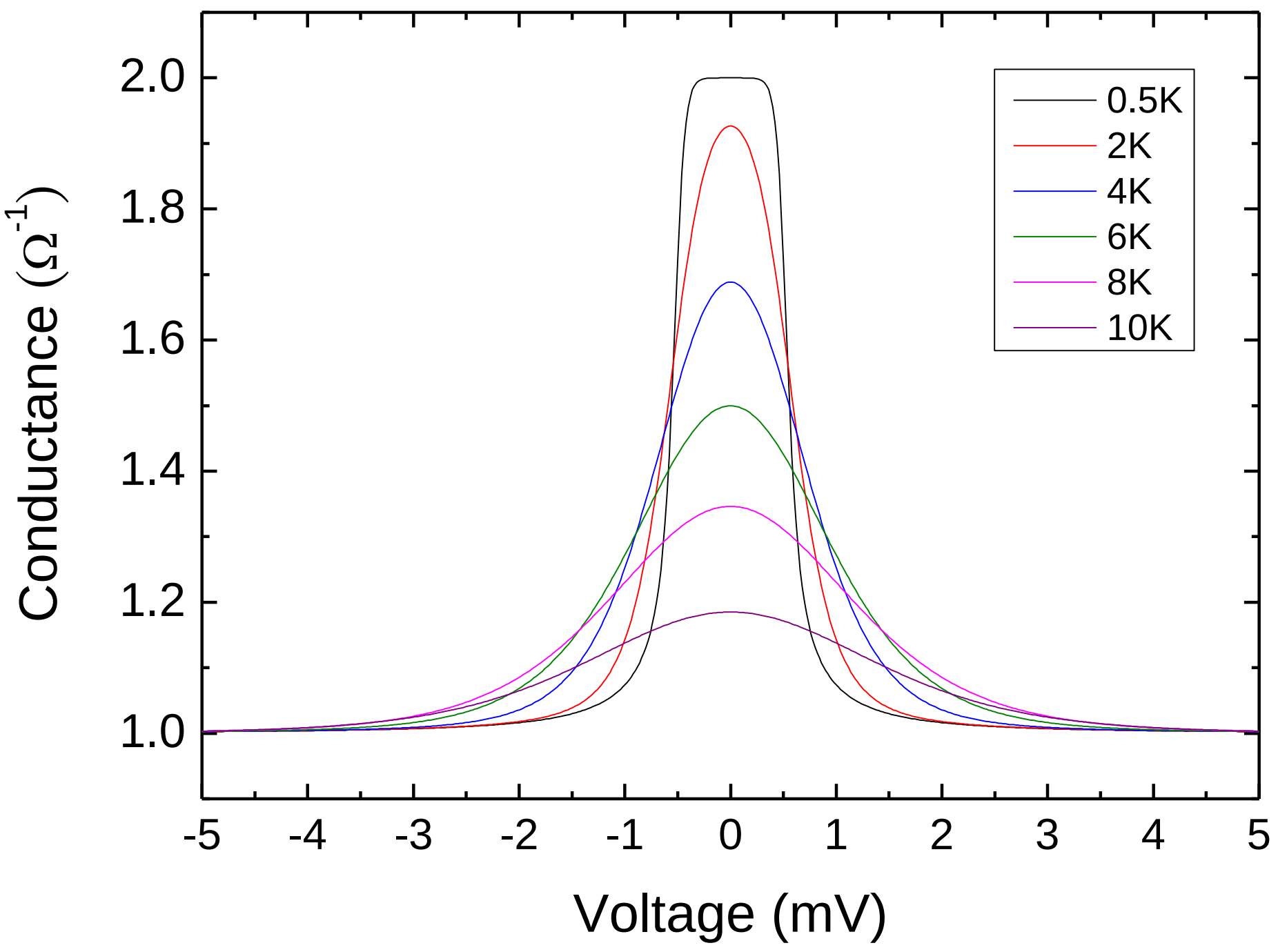

Fig. 7.

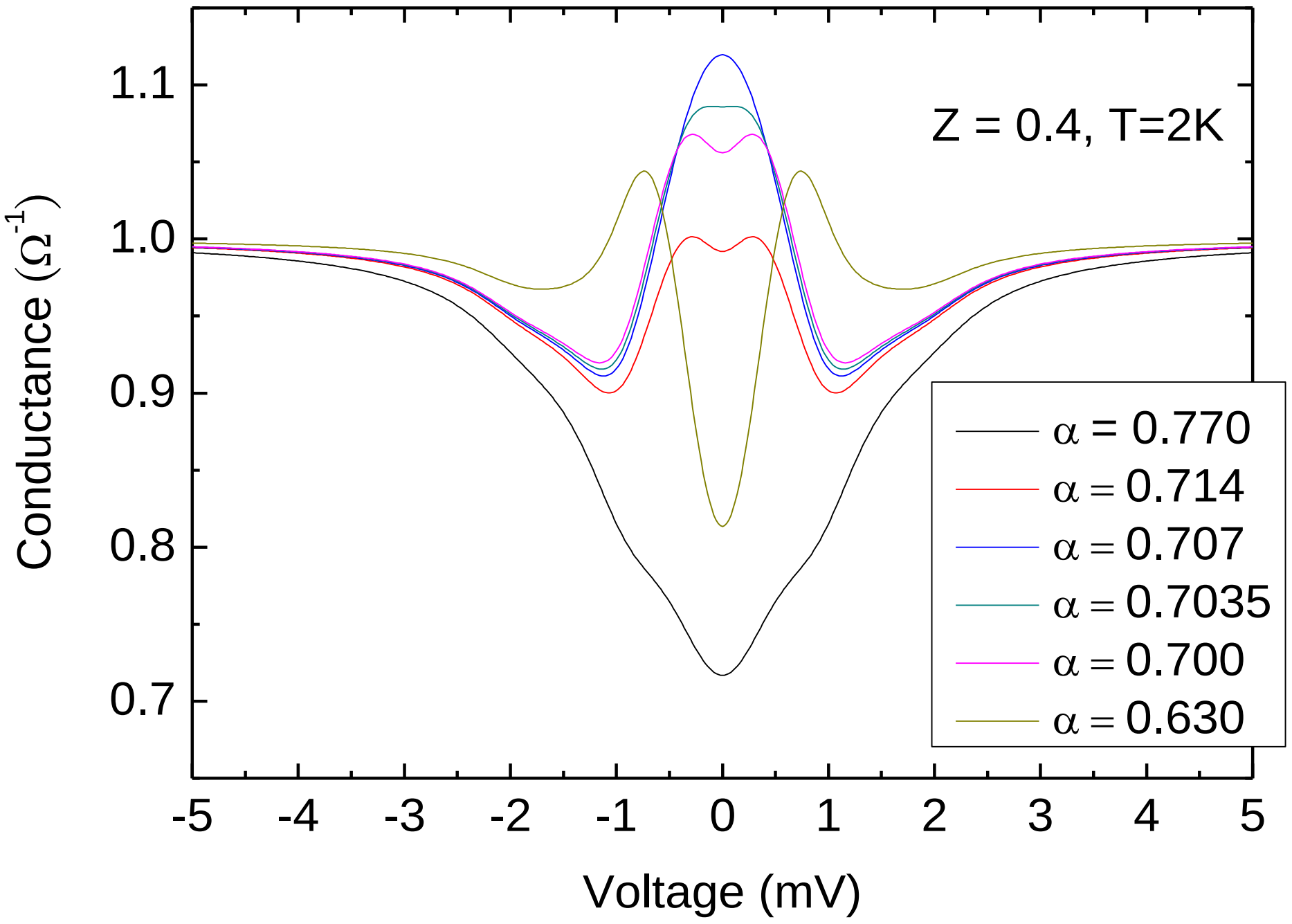